\documentclass[sigconf]{acmart}

\usepackage{enumitem}
\usepackage{calc}

\AtBeginDocument{%
  }

\copyrightyear{2024}
\acmYear{2024}
\setcopyright{acmlicensed}\acmConference[SIGIR-AP '24]{Proceedings of the 2024 Annual International ACM SIGIR Conference on Research and Development in Information Retrieval in the Asia Pacific Region}{December 9--12, 2024}{Tokyo, Japan}
\acmBooktitle{Proceedings of the 2024 Annual International ACM SIGIR Conference on Research and Development in Information Retrieval in the Asia Pacific Region (SIGIR-AP '24), December 9--12, 2024, Tokyo, Japan}
\acmDOI{10.1145/3673791.3698435}
\acmISBN{979-8-4007-0724-7/24/12}

\begin{document}

\title{R$^3$AG: First Workshop on Refined and Reliable \\ Retrieval Augmented Generation}


\author{Zihan Wang}
\affiliation{%
  \institution{University of Amsterdam}
  \city{Amsterdam}
  \country{The Netherlands}}
\email{zihanwang.sdu@gmail.com}

\author{Xuri Ge}
\affiliation{%
  \institution{University of Glasgow}
  \city{Glasgow}
  \country{United Kingdom}
}
\email{x.ge.2@research.gla.ac.uk}
\author{Joemon M. Jose}
\affiliation{%
  \institution{University of Glasgow}
  \city{Glasgow}
  \country{United Kingdom}}
\email{joemon.jose@glasgow.ac.uk}

\author{Haitao Yu}
\affiliation{%
  \institution{University of Tsukuba}
  \city{Tsukuba}
  \country{Japan}}
\email{yuhaitao@slis.tsukuba.ac.jp}

\author{Weizhi Ma}
\affiliation{%
  \institution{Tsinghua University}
  \city{Beijing}
  \country{China}}
\email{mawz12@hotmail.com}

\author{Zhaochun Ren}
\affiliation{%
  \institution{Leiden University}
  \city{Leiden}
  \country{The Netherlands}}
\email{z.ren@liacs.leidenuniv.nl}

\author{Xin Xin}
\affiliation{%
  \institution{Shandong University}
  \city{Shandong}
  \country{China}}
\email{xinxin@sdu.edu.cn}
\renewcommand{\shortauthors}{Zihan Wang et al.}
\newcommand{\todo}[1]{\textcolor{red}{[#1]}}

\begin{abstract}
  Retrieval-augmented generation (RAG) has gained wide attention as the key component 
  to improve generative models with external knowledge augmentation from information retrieval. It has shown great prominence in enhancing the functionality and performance of large language model (LLM)-based applications. 
  However, with the comprehensive application of RAG, more and more problems and limitations have been identified, thus urgently requiring further fundamental exploration to improve current RAG frameworks. 
  This workshop aims to explore in depth how to conduct refined and reliable RAG for downstream AI tasks.

  To this end, we propose to organize the first R$^3$AG workshop at SIGIR-AP 2024 to call for participants to re-examine and formulate the basic principles and practical implementation of refined and reliable RAG.
  The workshop serves as a platform for both academia and industry researchers to conduct discussions, share insights, and foster research to build the next generation of 
  RAG systems.
Participants will engage in discussions and presentations focusing on fundamental challenges, cutting-edge research, and  potential pathways to improve RAG.  
At the end of the workshop, we aim to have a clearer understanding of how to improve the reliability and applicability of RAG with more robust information retrieval and language generation. 
\end{abstract}





\maketitle
\section{BACKGROUND AND MOTIVATIONS}
RAG (Retrieval-Augmented Generation) has emerged as a new paradigm for using information retrieval (IR) to improve the generated response from large language models (LLMs).
On the one hand, traditional IR systems may encounter difficulties to handle more and more complex information seeking queries. On the other hand, LLM has shown notable natural language understanding capability while suffering from fictitious or inaccurate generation, also known as the hallucination problem. To this end, RAG has emerged to combine the best of IR and LLM generation. 
RAG has made significant progress in
improving response quality by first retrieving relevant knowledge from external knowledge base and then generating responses based on the knowledge retrieval.
RAG's advantages in handling information queries include but not limited to 
enhanced user experience, enriched information
return, improved response accuracy, and multi-round conversational search for complex queries. 
Through integrating IR and language generation, RAG has become the keystone for various AI applications.







Existing RAG techniques \cite{lewis2020retrieval,borgeaud2022improving} focused on enhancing language models by integrating additional textual knowledge from external knowledge database.
Transformer-based \cite{vaswani2017attention} language models have shown great promise in language generation, leading to notable LLMs like GPTs. 
LLMs owe their success to advanced architectures with billions of parameters, pre-trained on vast corpora from diverse sources, enabling remarkable generalization across various AI applications \cite{efeoglu2024retrieval,chen2024benchmarking,alkhalaf2024applying}.
However, LLMs also suffer from model hallucination \cite{huang2023survey} and difficulty in handling dynamic knowledge updates \cite{yao2023editing}. 
RAG alleviates the hallucination problem by providing LLMs with relevant knowledge using IR techniques to retrieve from external databases, achieving more accurate responses for knowledge-intensive generation. The decoupled database also supports more lightweight knowledge dynamic updates.
For example, \cite{li2024enhancing} integrates a RAG pipeline in an end-to-end generation system to improve the factual correctness of LLMs for domain-specific and time-sensitive queries. 



While RAG has achieved great advancement, we recognize that there still exists challenges to conduct refined and reliable RAG (R$^3$AG). 
A typical RAG pipeline often includes user intent comprehension, knowledge parsing, knowledge retrieval, and response generation. Each pipeline step has specific challenges and plays an essential role to accomplish user queries.
For example, how to understand user query intention under long and complex dialogue context; how to parse complex knowledge documents including tables and figures; how to conduct reliable knowledge retrieval; and how to refine the generated response. 
The workshop is expected to help researchers to conduct further investigation on R$^3$AG.


\noindent\textbf{Topic.} 
The topic of this workshop includes interesting points related to the RAG pipelines, i.e., user intent comprehension, knowledge parsing, knowledge retrieval, and response generation. The detailed topics are described in section \ref{sec:challenges}.



\noindent\textbf{Relevance.} 
On the one hand, generative LLMs are hot research topics in the IR communities. The capability of LLMs enriches the scope of IR and has changed the process of information seeking to a large extend. 
On the other hand, RAG is one of the most important applications of IR in LLMs. IR plays a new and essential role in LLM-based downstream tasks.
To this end, this workshop of R$^3$AG is highly relevant to the SIGIR-AP conference. 


\noindent\textbf{Motivation.} 
Recently, the potential of RAG has been verified in various AI applications. However, there still exists fundamental research challenges to further improve current RAG methods. 
SIGIR-AP is  one of the leading conferences with numerous recognized research works focusing on  IR-related research and applications.  
We believe that organizing the R$^3$AG workshop with SIGIR-AP now can (i) stimulate interesting research to meet complex information seeking demands; (ii) encourage the community to conduct in-depth research and practical applications on refined and reliable RAG; (iii) expand the impact of SIGIR-AP conferences and the workshop.


\section{FUNDAMENTAL CHALLENGES AND TOPICS}
\label{sec:challenges}
This year we will focus more on fundamental challenges in this field and expect thorough discussions during the R$^3$AG workshop.

\noindent \textbf{User Intent Comprehension.} 
User intent comprehension directly affects the following retrieval and final response generation. However, user intent comprehension could be challenging especially under long dialogue context. R$^3$AG is expected to help users clarify their information needs during the interaction process. Besides, R$^3$AG should have the capability to split complex queries into simple queries to accomplish user demands.
Related methods include query expansion, which introduces hypothetical answer generation from LLMs into the retrieval process to improve the retrieval relevance, query summarization \cite{edge2024local}, query rewrite \cite{mao2024rafe}, etc.


\noindent \textbf{Query \& Knowledge Encoding.}
The effectiveness of RAG depends heavily on the representation learning of both queries and knowledge. Mainstream methods use pre-trained feature extractors \cite{devlin2018bert,radford2021learning} to encode them, while there could be a misalignment between diverse knowledge and user queries. Currently, fine-turning based query-knowledge encoding plays an important role in RAG.
However, there still lacks sufficient exploration to conduct more effective query \& knowledge encoding for R$^3$AG, especially in terms of efficient fine-tuning to support knowledge updates.


\noindent \textbf{RAG for Complex Documents.} 
 RAG systems aim to deliver knowledge chunks to improve LLM generation. However, existing RAG methods could encounter difficulties to parse complex documents with embedded tables and figures. How to parse complex tabular knowledge is still an open research question. 
 Besides, the chunks division strategy also affects response generation. Too large chunks introduce irrelevant information while small chunks could lead to incomplete knowledge. 
 In addition, multi-hop document retrieval is also a challenge. The above three issues need to be further investigated to conduct R$^3$AG. 

\noindent \textbf{Reliable Retrieval for RAG.} 
The inclusion of noisy or contradictory information during retrieval can significantly impair the performance of existing RAG models. 
There has been growing interest in improving the resilience of RAG against harmful or counterfactual inputs, which is now considered a crucial performance indicator in recent studies~\citep{yoran2023making,yu2023chain,DBLP:conf/emnlp/BaekJKPH23}.
Meanwhile, current research~\citep{cuconasu2024power} reveals that including irrelevant documents can unexpectedly increase accuracy by more than 30\%, challenging the initial belief that it would degrade quality.
These findings inspire us to organize R$^3$AG for developing specialized strategies for effective retrieval and to emphasize the ongoing need for further investigation into RAG's robustness and reliability.


\noindent \textbf{Response Evaluation and Refinement.}  While RAG enhances LLMs by providing additional information, LLMs may still face challenges with unreliable or inaccurate response generation. These challenges may arise from incorrect information in the context~\citep{bian2023influence} or issues with hallucinations~\citep{DBLP:journals/tacl/AdlakhaBLMR24}. Unfortunately, there is a noticeable gap in understanding how these challenges impact the output quality of RAG, as well as in developing strategies for models to mitigate these issues and refine their responses.
As such, R$^3$AG is dedicated to comprehensively evaluating and enhancing the response quality of RAG-based LLMs across multiple dimensions, such as relevance, faithfulness, negative rejection, information integration, creative generation, and error correction.



\noindent \textbf{Multimodal R$^3$AG.} 
 Most current research focuses on textual RAG, despite the need for multimodality in many applications. Recent large-scale models like Flamingo \cite{alayrac2022flamingo}, and GPT-4 \cite{achiam2023gpt} show significant multimodal capability when scaled to tens of billions of parameters and trained on extensive multimodal corpora. These large multimodal models require RAG even more than unimodal textual LLMs for  external knowledge support. Therefore, the workshop encourage discussions regarding R$^3$AG for large multimodal models.


\section{PROGRAM SKETCH}
\subsection{Workshop Format}
The workshop is planned to be hosted for half a day, including 2 invited talks and 4 oral research talks. 
There are two encouraging types of invited talks:
(i) academic talks on fundamental research on the adaptability and reliability of RAG techniques; and (ii) industrial talks on the practice of designing or applying refined and reliable RAG techniques for real-world applications, including information retrieval and generative systems.

Each talk should be delivered as a slide-based lecture. A Q\&A session will follow the conclusion of each talk.

\subsection{Tentative Workshop Schedule}
The workshop schedule is planned with one half-day session:
\begin{itemize}
    \item[-] 9:00 - 9:10 Welcome \& opening 
    \item[-] 9:10 - 9:50 Academic invited talk 
    \item[-] 9:50 - 10:30 Industrial invited talk
    \item[-] 10:30 - 10:50 Coffee break
    \item[-] 10:50 - 11:40 Oral paper talks
    \item[-] 11:40 - 12:00 Panel discussion
\end{itemize}
\noindent
\textbf{Tentative Speakers.} The tentative speakers include academia
researchers, such as Dr. Xiangyu Zhao from the City University of Hong Kong, and Prof. Hideo Joho from University of Tsukuba and industrial staff, such as Dr. Alexandros Karatzoglou from Amazon.

\noindent
\textbf{Panel Discussion.} We are also considering hosting a panel discussion as the final part of the workshop. The decision to include this session will depend on the availability of panelists attending SIGIR-AP and the number of accepted research papers.

\noindent
\textbf{Contingency Plan.} 
To ensure a successful workshop, our contingency plan identifies key requirements (venue, speakers, materials) and assesses risks. We will have backup venues, alternate speakers, and a ready team for onsite support. A designated lead will oversee a communication plan, with specific roles for last-minute organizers to manage registration, technical support, and logistics.

\subsection{Selection Process}  
Each invited speaker should be highly esteemed within the community. Invitations should be agreed upon by all organizers without any disagreement.
The workshop accepts paper submissions via a standard peer-review process, expects 6$\sim$10 paper submissions, and accepts 3$\sim$4 papers.
Each submission is evaluated by at least two members of the program committee. The senior PCs or workshop organizers will make the final decision. Authors will receive detailed review comments and a notification letter.

\noindent
\textbf{Tentative Program Committee.} In addition to the current 7 organizers, we also plan to invite the following tentative PC members: (1) Dr. Chao Huang from University of Hong Kong, (2) Dr. Xiangyu Zhao from the City University of Hong Kong, (3) Dr. Andrew Yates from University of Amsterdam, and (4) Dr. Alexandros Karatzoglou from Amazon.

\subsection{Online Materials}
A website for the R$^3$AG workshop will be made available online. All relevant materials, including talk information, presentation slides, referred papers, speaker details, and related open-source projects, will be accessible on this website.

\subsection{Workshop Advertisement}
The R$^3$AG workshop will be promoted on various social media platforms to increase visibility and encourage paper submissions. These platforms include, but are not limited to, Twitter, Facebook, and WeChat. Additionally, the organizers will send personalized emails to further advertise the workshop.

\section{RELATED WORKSHOPS}
List of related workshops:
\begin{itemize}
    \item Information Retrieval's Role in RAG Systems \\ (SIGIR 2024\footnote{\url{https://coda.io/@rstless-group/ir-rag-sigir24}})
    \item Multimodal Representation and Retrieval \\ (SIGIR 2024\footnote{\url{https://mrr-workshop.github.io/}})
    \item Information Retrieval Meets Large Language Models \\ (WWW 2024\footnote{\url{https://irmeetsllm.github.io/}})
    \item Large Knowledge-Enhanced Models \\ (IJCAI 2024\footnote{\url{https://lkm2024.openkg.org/}})
    \item Knowledge Retrieval and Language Models \\ (ICML 2022\footnote{\url{https://knowledge-retrieval-workshop.github.io/}})
\end{itemize}
The Information Retrieval's Role in RAG Systems workshop at SIGIR (2024) explored retrieval's integral role in RAG frameworks. As multimodal LLMs and RAG gain traction, the Multimodal Representation and Retrieval workshop at SIGIR (2024) introduced the challenge of multimodal queries and documents. The Information Retrieval Meets LLMs workshop at WWW (2024) addressed issues like retrieval-generation collaboration and hallucination. The Large Knowledge-Enhanced Models workshop at IJCAI (2024) discussed integrating LLMs with symbolic knowledge, while the Knowledge Retrieval and Language Models workshop at ICML (2022) highlighted the limitations of knowledge retrieval.
R$^3$AG is the first workshop to focus on refined and reliable RAG techniques. It will feature invited talks, paper presentations, and the release of real datasets and code for future practice.

\section{ORGANIZERS INFORMATION}
\noindent
\textbf{Prof. Joemon M. Jose} is a Professor at the School of Computing Science, University of Glasgow, Scotland and a member of the Information Retrieval group. His research focuses on the following three themes: (i) Social Media Analytics; (ii) Multi-modal LLMs for information retrieval; (iii) Multimedia mining and search. He has published over 300 papers with more than 10,000 Google Scholar citations, and an h-index of 51. He leads the GAIR Lab investigating research issues related to the above themes. 
He has been serving as the program committee chair and member for numerous top international conferences (e.g., SIGIR, WWW, and ECIR). He also serves as a PC chair for SIGIR-AP 2024.
\vspace{0.5em}

\noindent
\textbf{Dr. Zhaochun Ren} 
is an Associate Professor at Leiden University. His research interests focus on  joint research in IR and natural language processing, with an emphasis on conversational information seeking, question-answering, and recommender systems. He aims to develop intelligent systems that can address complex user requests and solve core challenges in both information retrieval and natural language processing towards that goal. In addition to his academic experience, he worked on e-commerce search and recommendation at JD.com for 2+ years. He has co-organized workshops at SIGIR (2020), WSDM (2019, 2020), and ECIR 2025.
\vspace{0.5em}

\noindent
\textbf{Dr. Haitao Yu}  
 is a Tenured Associate Professor at University of Tsukuba and leading the Information Intelligence research group. His research focuses on Information Retrieval, Knowledge Graph, and Machine Learning. He has published numerous papers on top international conferences (e.g., WSDM, CIKM, SIGIR, WWW, ECIR, and AAAI) and journals (e.g., Information Processing and Management, and Information Retrieval Journal). He is the co-organizer of the NTCIR tasks of Temporalia-2 and AKG. He has been serving as the program committee member for numerous top international conferences (e.g., WSDM, CIKM, SIGIR, and ECIR). 
\vspace{0.5em}

\noindent
\textbf{Dr. Xin Xin} is a Tenure-Track Assistant Professor at the School of Computer Science and Technology of Shandong University. Before that, he earned his Ph.D. degree from the University of Glasgow. His current research interests include information retrieval, natural language processing, and reinforcement learning. He has published more than 40 papers in top conferences (e.g., WWW, SIGIR, ACL, WSDM) and journals (e.g., TOIS, TKDE), and received the Best Paper Honor Mention at WSDM 2024.  He has organized the DRL4IR workshop in SIGIR and KEIR workshop in ECIR.
\vspace{0.5em}

\noindent
\textbf{Dr. Weizhi Ma} is a Research Assistant Professor at the Institute for AI Industry Research (AIR) at Tsinghua University. He earned his B.E. and Ph.D. in the Department of Computer Science and Technology from Tsinghua University. Dr. Ma's research focuses on information retrieval, recommender systems, and LLM-powered agents. He has published over 70 papers in leading international conferences and journals, including TOIS, TKDE, SIGIR, KDD, AAAI, IJCAI, WSDM, CIKM, etc. His accolades include the Best Paper Honorable Mention Award at SIGIR 2020 and several other paper awards. In addition to his research, Dr. Ma is the Secretary of the Youth Working Committee of the Chinese Information Processing Society of China and serves as an Assistant Editor for ACM TOIS. He is a recipient of the Young Elite Scientists Sponsorship Program by CAST and the Shuimu Tsinghua Scholar Program.
\vspace{0.5em}

\noindent
\textbf{Zihan Wang} is a fourth-year PhD student at the University of Amsterdam (UvA). He has published over ten papers in prestigious conferences including KDD, SIGIR, CCS, and WSDM. He received the Best Student Paper Award at WSDM 2018. His current research focuses on information extraction, knowledge graph embedding, and the reliability of large language models.
\vspace{0.5em}

\noindent \textbf{Xuri Ge} is a fourth-year PhD student at the University of Glasgow (UofG). He has published over 15 papers in prestigious conferences and journals, including ACM MM, SIGIR, NeurIPS, IP\&M, CIKM, ICME and ACM TIST, etc. His current research focuses on information retrieval, efficient multimodal representation learning, and multimodal large language models. 
Xuri has organized 3DMM workshop in IEEE ICME2024.
\vspace{0.5em}

\bibliographystyle{ACM-Reference-Format}
\bibliography{sample-base}


\begin{thebibliography}{21}


\ifx \showCODEN    \undefined \def \showCODEN     #1{\unskip}     \fi
\ifx \showDOI      \undefined \def \showDOI       #1{#1}\fi
\ifx \showISBNx    \undefined \def \showISBNx     #1{\unskip}     \fi
\ifx \showISBNxiii \undefined \def \showISBNxiii  #1{\unskip}     \fi
\ifx \showISSN     \undefined \def \showISSN      #1{\unskip}     \fi
\ifx \showLCCN     \undefined \def \showLCCN      #1{\unskip}     \fi
\ifx \shownote     \undefined \def \shownote      #1{#1}          \fi
\ifx \showarticletitle \undefined \def \showarticletitle #1{#1}   \fi
\ifx \showURL      \undefined \def \showURL       {\relax}        \fi
\providecommand\bibfield[2]{#2}
\providecommand\bibinfo[2]{#2}
\providecommand\natexlab[1]{#1}
\providecommand\showeprint[2][]{arXiv:#2}

\bibitem[Achiam et~al\mbox{.}(2023)]%
        {achiam2023gpt}
\bibfield{author}{\bibinfo{person}{Josh Achiam}, \bibinfo{person}{Steven Adler}, \bibinfo{person}{Sandhini Agarwal}, \bibinfo{person}{Lama Ahmad}, \bibinfo{person}{Ilge Akkaya}, \bibinfo{person}{Florencia~Leoni Aleman}, \bibinfo{person}{Diogo Almeida}, \bibinfo{person}{Janko Altenschmidt}, \bibinfo{person}{Sam Altman}, \bibinfo{person}{Shyamal Anadkat}, {et~al\mbox{.}}} \bibinfo{year}{2023}\natexlab{}.
\newblock \showarticletitle{Gpt-4 technical report}.
\newblock \bibinfo{journal}{\emph{arXiv preprint arXiv:2303.08774}} (\bibinfo{year}{2023}).
\newblock


\bibitem[Adlakha et~al\mbox{.}(2024)]%
        {DBLP:journals/tacl/AdlakhaBLMR24}
\bibfield{author}{\bibinfo{person}{Vaibhav Adlakha}, \bibinfo{person}{Parishad BehnamGhader}, \bibinfo{person}{Xing~Han Lu}, \bibinfo{person}{Nicholas Meade}, {and} \bibinfo{person}{Siva Reddy}.} \bibinfo{year}{2024}\natexlab{}.
\newblock \showarticletitle{Evaluating Correctness and Faithfulness of Instruction-Following Models for Question Answering}.
\newblock \bibinfo{journal}{\emph{Trans. Assoc. Comput. Linguistics}}  \bibinfo{volume}{12} (\bibinfo{year}{2024}), \bibinfo{pages}{681--699}.
\newblock


\bibitem[Alayrac et~al\mbox{.}(2022)]%
        {alayrac2022flamingo}
\bibfield{author}{\bibinfo{person}{Jean{-}Baptiste Alayrac}, \bibinfo{person}{Jeff Donahue}, \bibinfo{person}{Pauline Luc}, \bibinfo{person}{Antoine Miech}, \bibinfo{person}{Iain Barr}, \bibinfo{person}{Yana Hasson}, \bibinfo{person}{Karel Lenc}, \bibinfo{person}{Arthur Mensch}, \bibinfo{person}{Katherine Millican}, \bibinfo{person}{Malcolm Reynolds}, \bibinfo{person}{Roman Ring}, \bibinfo{person}{Eliza Rutherford}, \bibinfo{person}{Serkan Cabi}, \bibinfo{person}{Tengda Han}, \bibinfo{person}{Zhitao Gong}, \bibinfo{person}{Sina Samangooei}, \bibinfo{person}{Marianne Monteiro}, \bibinfo{person}{Jacob~L. Menick}, \bibinfo{person}{Sebastian Borgeaud}, \bibinfo{person}{Andy Brock}, \bibinfo{person}{Aida Nematzadeh}, \bibinfo{person}{Sahand Sharifzadeh}, \bibinfo{person}{Mikolaj Binkowski}, \bibinfo{person}{Ricardo Barreira}, \bibinfo{person}{Oriol Vinyals}, \bibinfo{person}{Andrew Zisserman}, {and} \bibinfo{person}{Kar{\'{e}}n Simonyan}.} \bibinfo{year}{2022}\natexlab{}.
\newblock \showarticletitle{Flamingo: a Visual Language Model for Few-Shot Learning}. In \bibinfo{booktitle}{\emph{NeurIPS}}.
\newblock


\bibitem[Alkhalaf et~al\mbox{.}(2024)]%
        {alkhalaf2024applying}
\bibfield{author}{\bibinfo{person}{Mohammad Alkhalaf}, \bibinfo{person}{Ping Yu}, \bibinfo{person}{Mengyang Yin}, {and} \bibinfo{person}{Chao Deng}.} \bibinfo{year}{2024}\natexlab{}.
\newblock \showarticletitle{Applying generative AI with retrieval augmented generation to summarize and extract key clinical information from electronic health records}.
\newblock \bibinfo{journal}{\emph{Journal of Biomedical Informatics}} (\bibinfo{year}{2024}), \bibinfo{pages}{104662}.
\newblock


\bibitem[Baek et~al\mbox{.}(2023)]%
        {DBLP:conf/emnlp/BaekJKPH23}
\bibfield{author}{\bibinfo{person}{Jinheon Baek}, \bibinfo{person}{Soyeong Jeong}, \bibinfo{person}{Minki Kang}, \bibinfo{person}{Jong~C. Park}, {and} \bibinfo{person}{Sung~Ju Hwang}.} \bibinfo{year}{2023}\natexlab{}.
\newblock \showarticletitle{Knowledge-Augmented Language Model Verification}. In \bibinfo{booktitle}{\emph{EMNLP}}. \bibinfo{pages}{1720--1736}.
\newblock


\bibitem[Bian et~al\mbox{.}(2023)]%
        {bian2023influence}
\bibfield{author}{\bibinfo{person}{Ning Bian}, \bibinfo{person}{Hongyu Lin}, \bibinfo{person}{Peilin Liu}, \bibinfo{person}{Yaojie Lu}, \bibinfo{person}{Chunkang Zhang}, \bibinfo{person}{Ben He}, \bibinfo{person}{Xianpei Han}, {and} \bibinfo{person}{Le Sun}.} \bibinfo{year}{2023}\natexlab{}.
\newblock \showarticletitle{Influence of external information on large language models mirrors social cognitive patterns}.
\newblock \bibinfo{journal}{\emph{arXiv preprint arXiv:2305.04812}} (\bibinfo{year}{2023}).
\newblock


\bibitem[Borgeaud et~al\mbox{.}(2022)]%
        {borgeaud2022improving}
\bibfield{author}{\bibinfo{person}{Sebastian Borgeaud}, \bibinfo{person}{Arthur Mensch}, \bibinfo{person}{Jordan Hoffmann}, \bibinfo{person}{Trevor Cai}, \bibinfo{person}{Eliza Rutherford}, \bibinfo{person}{Katie Millican}, \bibinfo{person}{George~Bm Van Den~Driessche}, \bibinfo{person}{Jean-Baptiste Lespiau}, \bibinfo{person}{Bogdan Damoc}, \bibinfo{person}{Aidan Clark}, {et~al\mbox{.}}} \bibinfo{year}{2022}\natexlab{}.
\newblock \showarticletitle{Improving language models by retrieving from trillions of tokens}. In \bibinfo{booktitle}{\emph{ICML}}. \bibinfo{pages}{2206--2240}.
\newblock


\bibitem[Chen et~al\mbox{.}(2024)]%
        {chen2024benchmarking}
\bibfield{author}{\bibinfo{person}{Jiawei Chen}, \bibinfo{person}{Hongyu Lin}, \bibinfo{person}{Xianpei Han}, {and} \bibinfo{person}{Le Sun}.} \bibinfo{year}{2024}\natexlab{}.
\newblock \showarticletitle{Benchmarking large language models in retrieval-augmented generation}. In \bibinfo{booktitle}{\emph{AAAI}}, Vol.~\bibinfo{volume}{38}. \bibinfo{pages}{17754--17762}.
\newblock


\bibitem[Cuconasu et~al\mbox{.}(2024)]%
        {cuconasu2024power}
\bibfield{author}{\bibinfo{person}{Florin Cuconasu}, \bibinfo{person}{Giovanni Trappolini}, \bibinfo{person}{Federico Siciliano}, \bibinfo{person}{Simone Filice}, \bibinfo{person}{Cesare Campagnano}, \bibinfo{person}{Yoelle Maarek}, \bibinfo{person}{Nicola Tonellotto}, {and} \bibinfo{person}{Fabrizio Silvestri}.} \bibinfo{year}{2024}\natexlab{}.
\newblock \showarticletitle{The power of noise: Redefining retrieval for rag systems}.
\newblock \bibinfo{journal}{\emph{arXiv preprint arXiv:2401.14887}} (\bibinfo{year}{2024}).
\newblock


\bibitem[Devlin et~al\mbox{.}(2018)]%
        {devlin2018bert}
\bibfield{author}{\bibinfo{person}{Jacob Devlin}, \bibinfo{person}{Ming-Wei Chang}, \bibinfo{person}{Kenton Lee}, {and} \bibinfo{person}{Kristina Toutanova}.} \bibinfo{year}{2018}\natexlab{}.
\newblock \showarticletitle{Bert: Pre-training of deep bidirectional transformers for language understanding}.
\newblock \bibinfo{journal}{\emph{arXiv preprint arXiv:1810.04805}} (\bibinfo{year}{2018}).
\newblock


\bibitem[Edge et~al\mbox{.}(2024)]%
        {edge2024local}
\bibfield{author}{\bibinfo{person}{Darren Edge}, \bibinfo{person}{Ha Trinh}, \bibinfo{person}{Newman Cheng}, \bibinfo{person}{Joshua Bradley}, \bibinfo{person}{Alex Chao}, \bibinfo{person}{Apurva Mody}, \bibinfo{person}{Steven Truitt}, {and} \bibinfo{person}{Jonathan Larson}.} \bibinfo{year}{2024}\natexlab{}.
\newblock \showarticletitle{From local to global: A graph rag approach to query-focused summarization}.
\newblock \bibinfo{journal}{\emph{arXiv preprint arXiv:2404.16130}} (\bibinfo{year}{2024}).
\newblock


\bibitem[Efeoglu and Paschke(2024)]%
        {efeoglu2024retrieval}
\bibfield{author}{\bibinfo{person}{Sefika Efeoglu} {and} \bibinfo{person}{Adrian Paschke}.} \bibinfo{year}{2024}\natexlab{}.
\newblock \showarticletitle{Retrieval-Augmented Generation-based Relation Extraction}.
\newblock \bibinfo{journal}{\emph{arXiv preprint arXiv:2404.13397}} (\bibinfo{year}{2024}).
\newblock


\bibitem[Huang et~al\mbox{.}(2023)]%
        {huang2023survey}
\bibfield{author}{\bibinfo{person}{Lei Huang}, \bibinfo{person}{Weijiang Yu}, \bibinfo{person}{Weitao Ma}, \bibinfo{person}{Weihong Zhong}, \bibinfo{person}{Zhangyin Feng}, \bibinfo{person}{Haotian Wang}, \bibinfo{person}{Qianglong Chen}, \bibinfo{person}{Weihua Peng}, \bibinfo{person}{Xiaocheng Feng}, \bibinfo{person}{Bing Qin}, {et~al\mbox{.}}} \bibinfo{year}{2023}\natexlab{}.
\newblock \showarticletitle{A survey on hallucination in large language models: Principles, taxonomy, challenges, and open questions}.
\newblock \bibinfo{journal}{\emph{arXiv preprint arXiv:2311.05232}} (\bibinfo{year}{2023}).
\newblock


\bibitem[Lewis et~al\mbox{.}(2020)]%
        {lewis2020retrieval}
\bibfield{author}{\bibinfo{person}{Patrick S.~H. Lewis}, \bibinfo{person}{Ethan Perez}, \bibinfo{person}{Aleksandra Piktus}, \bibinfo{person}{Fabio Petroni}, \bibinfo{person}{Vladimir Karpukhin}, \bibinfo{person}{Naman Goyal}, \bibinfo{person}{Heinrich K{\"{u}}ttler}, \bibinfo{person}{Mike Lewis}, \bibinfo{person}{Wen{-}tau Yih}, \bibinfo{person}{Tim Rockt{\"{a}}schel}, \bibinfo{person}{Sebastian Riedel}, {and} \bibinfo{person}{Douwe Kiela}.} \bibinfo{year}{2020}\natexlab{}.
\newblock \showarticletitle{Retrieval-Augmented Generation for Knowledge-Intensive {NLP} Tasks}. In \bibinfo{booktitle}{\emph{NeurIPS}}.
\newblock


\bibitem[Li et~al\mbox{.}(2024)]%
        {li2024enhancing}
\bibfield{author}{\bibinfo{person}{Jiarui Li}, \bibinfo{person}{Ye Yuan}, {and} \bibinfo{person}{Zehua Zhang}.} \bibinfo{year}{2024}\natexlab{}.
\newblock \showarticletitle{Enhancing llm factual accuracy with rag to counter hallucinations: A case study on domain-specific queries in private knowledge-bases}.
\newblock \bibinfo{journal}{\emph{arXiv preprint arXiv:2403.10446}} (\bibinfo{year}{2024}).
\newblock


\bibitem[Mao et~al\mbox{.}(2024)]%
        {mao2024rafe}
\bibfield{author}{\bibinfo{person}{Shengyu Mao}, \bibinfo{person}{Yong Jiang}, \bibinfo{person}{Boli Chen}, \bibinfo{person}{Xiao Li}, \bibinfo{person}{Peng Wang}, \bibinfo{person}{Xinyu Wang}, \bibinfo{person}{Pengjun Xie}, \bibinfo{person}{Fei Huang}, \bibinfo{person}{Huajun Chen}, {and} \bibinfo{person}{Ningyu Zhang}.} \bibinfo{year}{2024}\natexlab{}.
\newblock \showarticletitle{RaFe: Ranking Feedback Improves Query Rewriting for RAG}.
\newblock \bibinfo{journal}{\emph{arXiv preprint arXiv:2405.14431}} (\bibinfo{year}{2024}).
\newblock


\bibitem[Radford et~al\mbox{.}(2021)]%
        {radford2021learning}
\bibfield{author}{\bibinfo{person}{Alec Radford}, \bibinfo{person}{Jong~Wook Kim}, \bibinfo{person}{Chris Hallacy}, \bibinfo{person}{Aditya Ramesh}, \bibinfo{person}{Gabriel Goh}, \bibinfo{person}{Sandhini Agarwal}, \bibinfo{person}{Girish Sastry}, \bibinfo{person}{Amanda Askell}, \bibinfo{person}{Pamela Mishkin}, \bibinfo{person}{Jack Clark}, {et~al\mbox{.}}} \bibinfo{year}{2021}\natexlab{}.
\newblock \showarticletitle{Learning transferable visual models from natural language supervision}. In \bibinfo{booktitle}{\emph{ICML}}. \bibinfo{pages}{8748--8763}.
\newblock


\bibitem[Vaswani et~al\mbox{.}(2017)]%
        {vaswani2017attention}
\bibfield{author}{\bibinfo{person}{Ashish Vaswani}, \bibinfo{person}{Noam Shazeer}, \bibinfo{person}{Niki Parmar}, \bibinfo{person}{Jakob Uszkoreit}, \bibinfo{person}{Llion Jones}, \bibinfo{person}{Aidan~N Gomez}, \bibinfo{person}{{\L}ukasz Kaiser}, {and} \bibinfo{person}{Illia Polosukhin}.} \bibinfo{year}{2017}\natexlab{}.
\newblock \showarticletitle{Attention is all you need}. In \bibinfo{booktitle}{\emph{NeurIPS}}. \bibinfo{pages}{5998--6008}.
\newblock


\bibitem[Yao et~al\mbox{.}(2023)]%
        {yao2023editing}
\bibfield{author}{\bibinfo{person}{Yunzhi Yao}, \bibinfo{person}{Peng Wang}, \bibinfo{person}{Bozhong Tian}, \bibinfo{person}{Siyuan Cheng}, \bibinfo{person}{Zhoubo Li}, \bibinfo{person}{Shumin Deng}, \bibinfo{person}{Huajun Chen}, {and} \bibinfo{person}{Ningyu Zhang}.} \bibinfo{year}{2023}\natexlab{}.
\newblock \showarticletitle{Editing large language models: Problems, methods, and opportunities}. In \bibinfo{booktitle}{\emph{EMNLP}}.
\newblock


\bibitem[Yoran et~al\mbox{.}(2023)]%
        {yoran2023making}
\bibfield{author}{\bibinfo{person}{Ori Yoran}, \bibinfo{person}{Tomer Wolfson}, \bibinfo{person}{Ori Ram}, {and} \bibinfo{person}{Jonathan Berant}.} \bibinfo{year}{2023}\natexlab{}.
\newblock \showarticletitle{Making retrieval-augmented language models robust to irrelevant context}.
\newblock \bibinfo{journal}{\emph{arXiv preprint arXiv:2310.01558}} (\bibinfo{year}{2023}).
\newblock


\bibitem[Yu et~al\mbox{.}(2023)]%
        {yu2023chain}
\bibfield{author}{\bibinfo{person}{Wenhao Yu}, \bibinfo{person}{Hongming Zhang}, \bibinfo{person}{Xiaoman Pan}, \bibinfo{person}{Kaixin Ma}, \bibinfo{person}{Hongwei Wang}, {and} \bibinfo{person}{Dong Yu}.} \bibinfo{year}{2023}\natexlab{}.
\newblock \showarticletitle{Chain-of-note: Enhancing robustness in retrieval-augmented language models}.
\newblock \bibinfo{journal}{\emph{arXiv preprint arXiv:2311.09210}} (\bibinfo{year}{2023}).
\newblock


\end{thebibliography}

\end{document}